\documentclass[doublecol]{epl2}

\title{Antisite Domains in Double Perovskite Ferromagnets: \\
Impact on Magnetotransport and Half-metallicity} 

\shorttitle{Antisites in double perovskites}

\author{Viveka Nand Singh \and Pinaki Majumdar}
\shortauthor{Viveka Nand Singh and Pinaki Majumdar}

\institute{
\inst{} Harish-Chandra  Research Institute,
Chhatnag Road, Jhusi, Allahabad 211019, India
}


\abstract{
Several double perovskite materials of the form A$_{2}$BB'O$_{6}$ 
exhibit high ferromagnetic T$_{c}$, and significant low field 
magnetoresistance. They are also a candidate source of spin polarized
electrons. The potential usefulness of these materials is, however,
frustrated by mislocation of the B and B' ions, which do not
organise themselves in the ideal alternating structure. The result is
a strong dependence of physical properties on preparative conditions, reducing 
the magnetization and destroying the half-metallicity.  We provide the first
results on the impact of \textit{spatially correlated} antisite disorder, 
as observed experimentally, on the ferromagnetic double perovskites. The
antisite domains suppress magnetism and half-metallicity, as expected, 
but lead to a dramatic enhancement of  the low field magnetoresistance. 
}

\begin{document}

\maketitle

Double perovskite (DP) materials, 
of the form A$_{2}$BB'O$_{6}$, are of interest
on account of their unusual magnetic and transport 
properties \cite{dp-rev}. 
In particular, some double perovskites, 
{\it e.g.}, Sr$_{2}$FeMoO$_{6}$,
show high ferromagnetic $T_c, \sim$ 420K, half-metallic behavior,  
and large low field magnetoresistance \cite{nat-kob,tom-cryst}.  
Other double perovskites,
{\it e.g.}, Sr$_{2}$FeWO$_{6}$, 
are antiferromagnetic insulators \cite{ins-pap}, 
while some show unusual dielectric 
properties \cite{lanimn-dd}.

Usually one of the ions,  B, say, is magnetic and the 
other (B') is non-magnetic.
Magnetic ordering in these materials arises from a 
combination of strong electron-spin coupling on the B ion
and electron delocalisation on the B-O-B' network 
\cite{theor-abin-dd,theor-millis,ps-pm-scr}. 
The magnetic order, however, is also strongly affected by the
{\it local ordering} of the B and B' 
ions \cite{garcia-hern,huang-asd1,huang-asd2,nav-asd1,nav-asd2,dd-gb-mr,asd-theor1,asd-theor2}. 
A B-O-B arrangement,
for example, favours antiferromagnetic locking of the
two B moments rather than parallel alignment.
Most real materials have some degree  of such
`antisite disorder' (ASD) with mislocated B and B' 
ions. 
The promise of rich functionality in the DP's
remains unfulfilled due to
this inevitable B, B' mislocation. 

Let us summarise the key observations on Sr$_{2}$FeMoO$_{6}$ (SFMO), 
the
prototype DP, to establish the tasks for a theory of 
antisite disorder 
in these materials. (i)~{\it Nature of defects:}
there is clear evidence now that B-B' mislocations 
are not random but spatially correlated \cite{asaka-asd-dom,dd-asd-dom}. 
While ASD
suppresses long range structural order, 
electron microscopy \cite{asaka-asd-dom} and XAFS \cite{dd-asd-dom}
reveal that  a high degree of short range 
order survives. (ii)~{\it  Magnetism:} the
ASD brings into play a B-B  AF coupling, via B-O-B links,
that makes neighbouring FM regions antiparallel. 
This reduces the bulk moment from $M_{max} \sim 4\mu_B$ 
expected in `ordered' SFMO.
The $T_c$ is {\it not} significantly affected by
ASD, till very large disorder. The magnetic effects of ASD 
are similar in both single crystals 
and polycrystals \cite{nat-kob,tom-cryst,huang-asd1}. 
(iii)~{\it  Transport:} 
resistivity 
results are widely different between
single crystals and polycrystals. Single crystals \cite{tom-cryst}
show residual  resistivity $ \rho \sim 0.1$m$\Omega$cm, 
and metallic behaviour, $d\rho/dT >0$.
The magnetoresistance (MR) is weak, $< 10 \%$ at low
temperature at a field of $5$~Tesla.
Unfortunately, single crystal data
with systematic variation of ASD is not
available.
Polycrystalline samples have been 
studied \cite{huang-asd1,huang-asd2,nav-asd1,nav-asd2,dd-gb-mr,niebel-asd} 
for a wide range of ASD {\it but} the transport in
these materials is affected also by grain boundary (GB)
resistance. 
The residual resistivity in
these samples \cite{huang-asd1}  range from 
 $\sim 0.5$m$\Omega$cm for low
ASD ($M/M_{max} \sim 1.0$) to $\sim 10$m$\Omega$cm
at high ASD  ($M/M_{max} \sim 0.5$). 
`Ordered' polycrystals show $d\rho/dT >0$, while the
less ordered ones  \cite{huang-asd1,huang-asd2} 
show $d\rho/dT <0$.
The MR can be large, $\sim 40\%$ at
low temperature and 5 Tesla \cite{huang-asd2}, and seems to be dominated 
by grain boundary effects \cite{dd-gb-mr,niebel-asd}. 
Some results indicate a decrease \cite{huang-asd2} in MR with 
increasing ASD,  others show an increase \cite{nav-asd1}.
The effects of antisite disorder and grain
boundaries on MR have not been deconvolved yet. 

In this paper we {\it build in} the correlated 
antisite disorder and study its impact on magnetism and 
{\it intrinsic transport} in the double perovskites.
We ignore the effect of grain boundaries, admittedly
important in polycrystals, but uncover a
metal-insulator transition and large MR that can result
from antisite disorder itself. 

Our main results, based on 
large lattices in two dimensions (2D), are the
following.
$(i)$~Correlated antisite disorder leads to structural antiphase 
boundaries (APB) that also act
as magnetic domain walls (MDW) at low temperature.
$(ii)$~Growing ASD suppresses the
saturation magnetisation, leaves the $T_c$ mostly unaffected,
but leads to a metal-insulator transition in
the electronic ground state.
$(iii)$~The `insulating' samples have huge low field MR
 at low temperature since 
domain rotation `unblocks' conduction paths.
$(iv)$~The conduction  spin polarisation roughly
follows the core spin magnetisation in its disorder and temperature
dependence.
Overall, moderate ASD seems to be {\it beneficial}, since it
enhances the MR without destroying the half-metallicity or
suppressing the $T_c$.

\section{Model and method}
 
The ASD arises from an annealing 
process, and is spatially correlated 
\cite{asaka-asd-dom,dd-asd-dom}
rather than random. 
Generating antisite disordered configurations that 
incorporate the detailed B, B' chemistry and
capture specific preparative conditions is difficult. 
However, one can model the inherent ordering
tendency in the B, B' system, and frustrate it via
poor annealing to mimic the situation in the material.
Using the structural motifs that emerge  
\cite{ps-pm-asd} we can then solve the 
coupled electronic-magnetic problem
through a real space technique.

We use a binary variable $\eta_i$ to encode
information about atomic positions. We set $\eta_i =1$ for B sites
and $\eta_i =0$ for B' sites. In a non disordered DP the $\eta_i$ will
alternate along each axis. We will consider spatially correlated
disordered configurations of $\eta$, discussed later, and for
any specified $\{\eta\}$ background the electronic-magnetic model
has the form:
\begin{eqnarray}
H &=&
\epsilon_{B} \sum_{i\sigma} \eta_{i} f_{i\sigma}^{\dagger}f_{i\sigma}+
\epsilon_{B'} \sum_{i\sigma}\left(1-\eta_{i}\right)  
m_{i\sigma}^{\dagger}m_{i\sigma}\nonumber\\
&& + H_{kin} \{ \eta\} 
+J\sum_{i \alpha \beta}\eta_{i} {\bf S}_{i}\cdot 
f_{i\alpha}^{\dagger}\overrightarrow{\sigma}_{\alpha\beta}f_{i\beta}\nonumber\\
&&+J_{AF}\sum_{\left\langle i,j\right\rangle }\eta_{i}
\eta_{j} {\bf S}_i\cdot{\bf S}_j
- h\sum_i S_{iz}
\end{eqnarray}

Here $f$ is the electron operator corresponding to the magnetic 
B site and $m$ is that of the non-magnetic 
B'  site. $\epsilon_B$ and 
$\epsilon_{B'}$ are onsite energies, 
at the B and B' sites respectively.
$\epsilon_{B} - \epsilon_{B'} $ 
is a  `charge transfer' energy.
$H_{kin}$ is the electron hopping term:
$-t_1\sum_{{\langle i,j \rangle} \sigma}\eta_i \eta_j
f_{i\sigma}^{\dagger}f_{j\sigma} $ 
$ -t_2\sum_{ \langle i,j \rangle \sigma }
( 1-\eta_i) (1-\eta_j) 
m_{i\sigma}^{\dagger}m_{j\sigma} $ 
$ -t_3\sum_{{ \langle i,j \rangle} \sigma}
( \eta_i+\eta_j-2\eta_i \eta_j)
( f_{i\sigma}^{\dagger}m_{j\sigma}+h.c)$.
The $t$'s are all nearest neighbour hopping amplitudes,
and for simplicity we set $t_1=t_2=t_3=t$ here.
${\bf S}_i$ is
the core spin on the site ${\bf R}_i$, and
$\vert {\bf S}_i \vert =1$. 
$J$ is the Hund's
coupling on the B sites, and we use $J/t \gg 1$. When the `up spin' core levels
are fully filled, as for Fe in SFMO, the conduction electron is forced
to be {\it antiparallel} to the core spin. We have used $J>0$ to
model this situation. For the present study we have set \cite{delta-ref}
the {\it effective } charge transfer energy 
$\epsilon_B -J/2 - \epsilon_{B'} =0$.

If two magnetic atoms are on neighbouring sites,
we also have to consider the {\it antiferromagnetic} (AF) coupling 
between them. Much of the physics of these
materials arises from the competition between delocalisation
driven ferromagnetic exchange and this B-B superexchange.
We set $J_{AF}S^2/t = 0.08$, based on the $T_N$ scale in SrFeO$_3$.
$h$ is an applied magnetic field in the ${\hat z}$ direction

The model above treats the variables $\eta_i$ as given.
What governs the distribution  
of ionic positions $\{\eta_i\}$?
The effective interaction between ions has a `direct' component
(Lennard-Jones, say) that we label
$V_{ion}$. There is also the {\it indirect} interaction via 
the electronic-magnetic degrees of freedom in $H \{\eta_i\}$. 
The total Hamiltonian, including all degrees of freedom, would be:
$H_{tot}
= H \{\eta_i\} + V_{ion}\{\eta_i\}$.
To obtain the effective interaction among the $\{\eta_i\}$
one should trace out the electronic and magnetic
variables. So, the effective potential $V_{eff} \{ \eta \}$
controlling the positional order is
$
V_{eff} \{ \eta \}
= -(1/\beta) log\int {\cal D} {\bf S}_i 
Tr_{\{f,m\}} e^{-\beta H_{tot}}.
$
Unfortunately 
there is limited information about $V_{ion}$, and 
the trace is computationally demanding.
Thankfully, the experiments themselves suggest a way ahead.

Firstly, the structural degrees of freedom `freeze' at a temperature 
$T_{ord} \sim 1000$K $\gg T_c \sim 400$K, where $T_c$ 
is the magnetic ordering temperature, so we need not
bother about the `feedback' of magnetic and electronic 
degrees of freedom on the $\{\eta_i\}$ ordering. 
Secondly,  
the tendency is to order
 into an alternating pattern, frustrated by
short annealing. So, as a first attempt, we can try out a simple
short range model with the same tendency. Concretely, we use
a `lattice gas' model
$
V_{eff} \{ \eta \} = -V \sum_{\langle ij\rangle} \eta_i (1 - \eta_j)
$
with  $V > 0$ being a measure of the ordering tendency. The
ground state in this model would correspond to $\eta =1,0,1,0,..$
along each axis, {\it i.e}, B, B', B, B',..
This approach tries to incorporate the effect of complex interactions 
between the A, B, B'  and O ions (as also the electrons) 
into a single parameter $V$.

We explored \cite{ps-pm-asd}
 the consequences of different annealing
protocols on this model, to examine
the consequences of imperfect annealing.
The lattice gas model has a finite temperature transition in both
2D and 3D. We work in 2D 
since it allows access to large sizes
and is easier to visualise.

\vfill
\begin{figure}[t]
\centerline{
\includegraphics[width=7.5cm,height=7.5cm,angle=0]{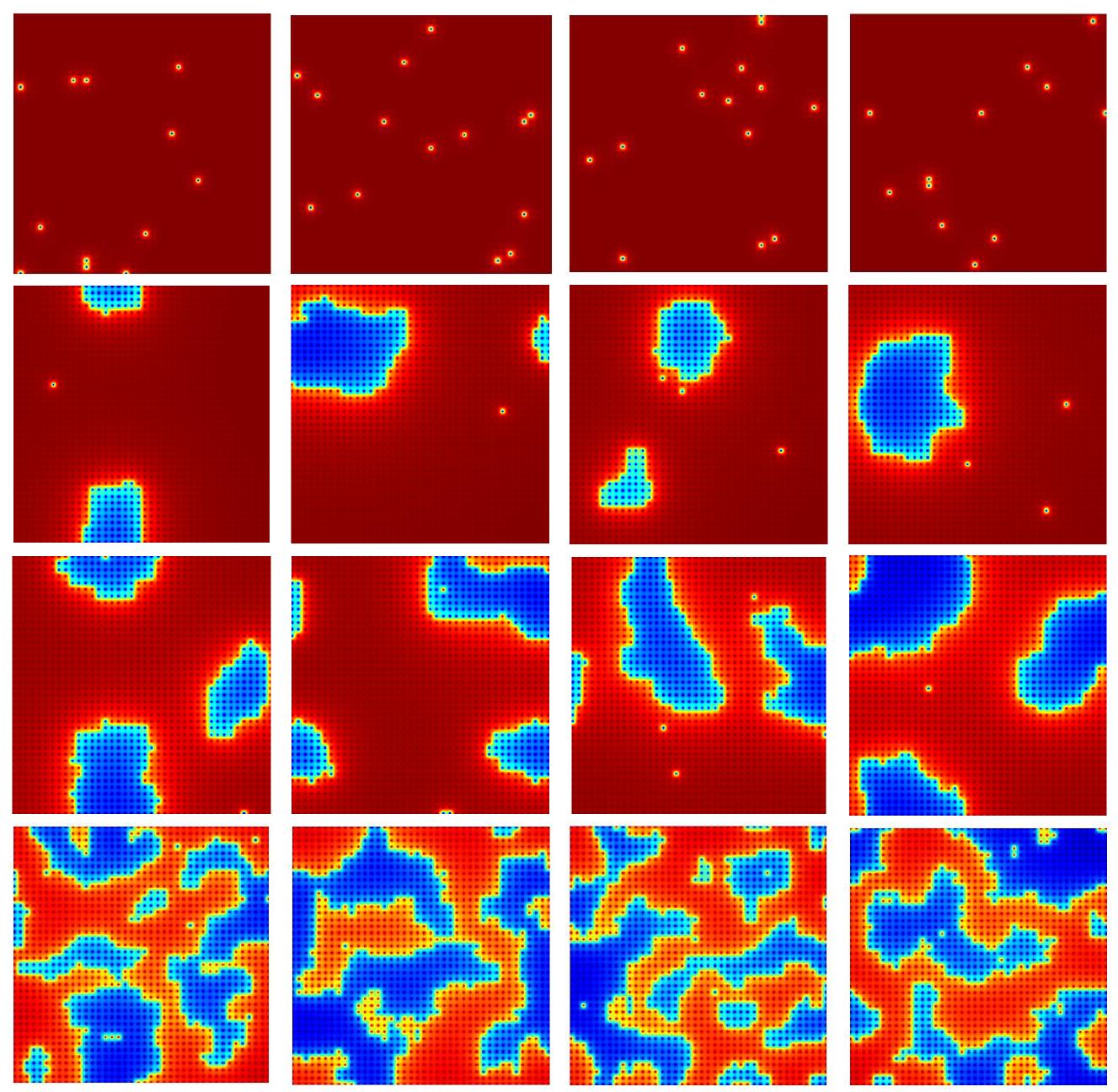}
}
\caption{Colour online: We show four families of antisite disordered
configurations (top to bottom)  generated via successively poorer
annealing of the lattice gas model. We plot 
$ (\eta_i - 1/2) e^{i \pi  (x_i + y_i)}$. For a perfectly ordered structure
$g({\bf R}_i)$ is constant. The patterns along a row are
different
realisations of ASD within each family. The average 
structural order parameter
(see text) has values $S=0.98,~0.76,~0.50$ and $0.08$ as we move from
top to bottom. Lattice size $40 \times 40$.}
\end{figure}
\vfill

One can variously characterise the B-B' structures that emerge. 
We
use the following simple indicators, to keep a close correspondence 
with the experimental work. 
$(i)$ The fraction of
B (or B') atoms that are  on the wrong sublattice, call this $x$,
and the structural `order parameter'  $S=1-2x$.
$(ii)$~The degree of short range order, characterised by the probability,
$p$, of 
having nearest neighbour pairs that are B-B'. 
$(iii)$~The correlation length $\xi$ in these structures,
computed from the width of the ordering peak.

\begin{figure}[t]
\centerline{
\includegraphics[width=4.4cm,height=5.0cm,angle=0]{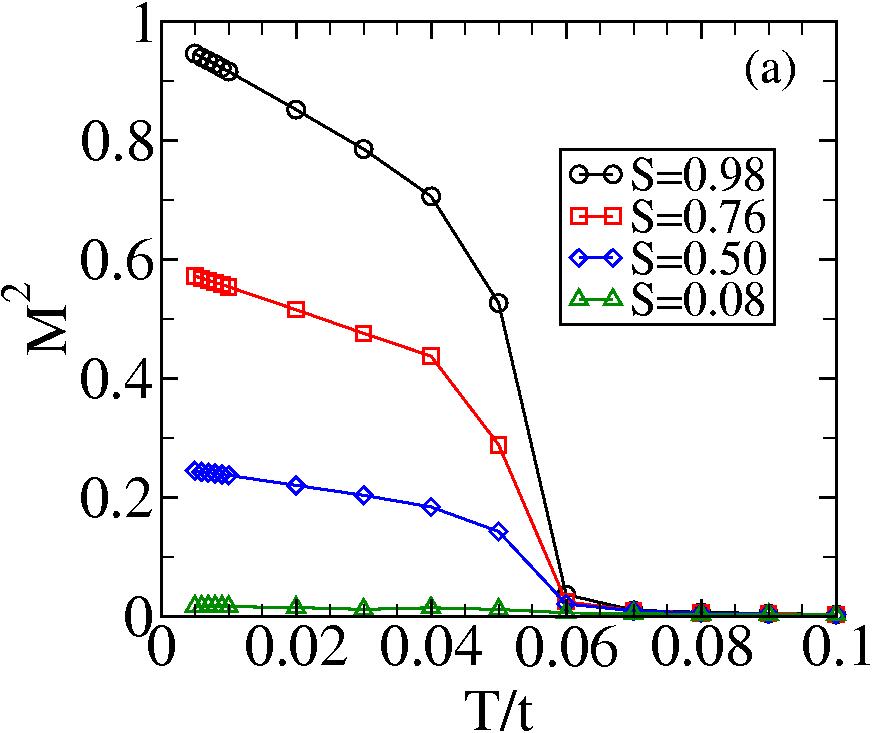}
\hspace{-.3cm}
\includegraphics[width=4.4cm,height=5.0cm,angle=0]{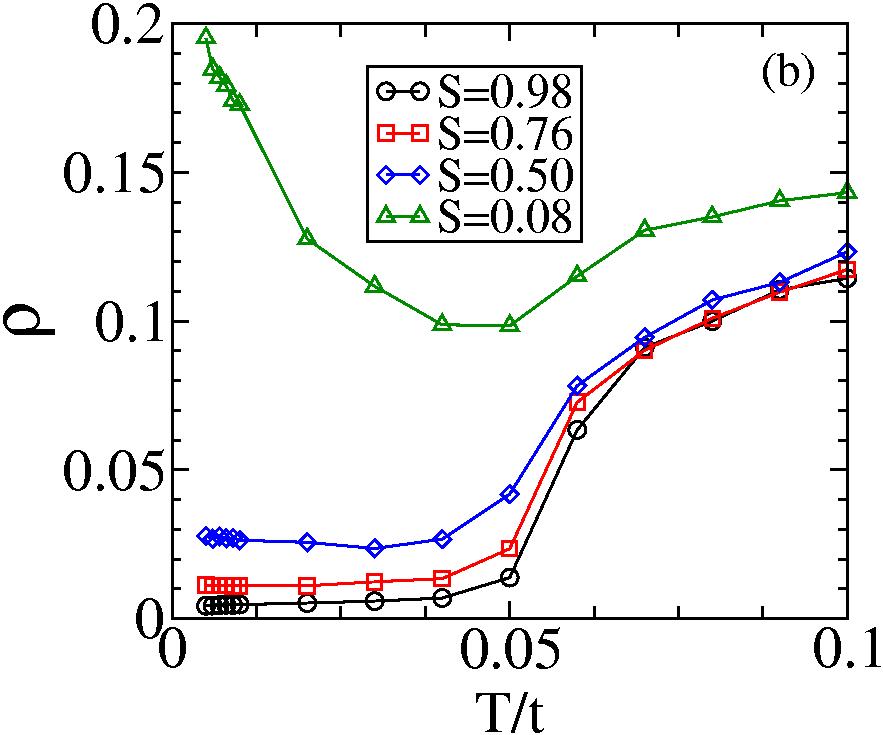}
}
\caption{Colour online: (a).~The ferromagnetic peak in the structure factor,
$M^2$, where $M$ is the magnetisation, for different degrees of
antisite disorder. (b)~The `d.c resistivity', $\rho(T)$ for
varying antisite disorder.
Results are on a $40 \times 40$ lattice, averaged thermally and 
over 10 copies of disorder.}
\end{figure}

\begin{figure}[t]
\centerline{
\includegraphics[width=7.8cm,height=7.8cm,angle=0]{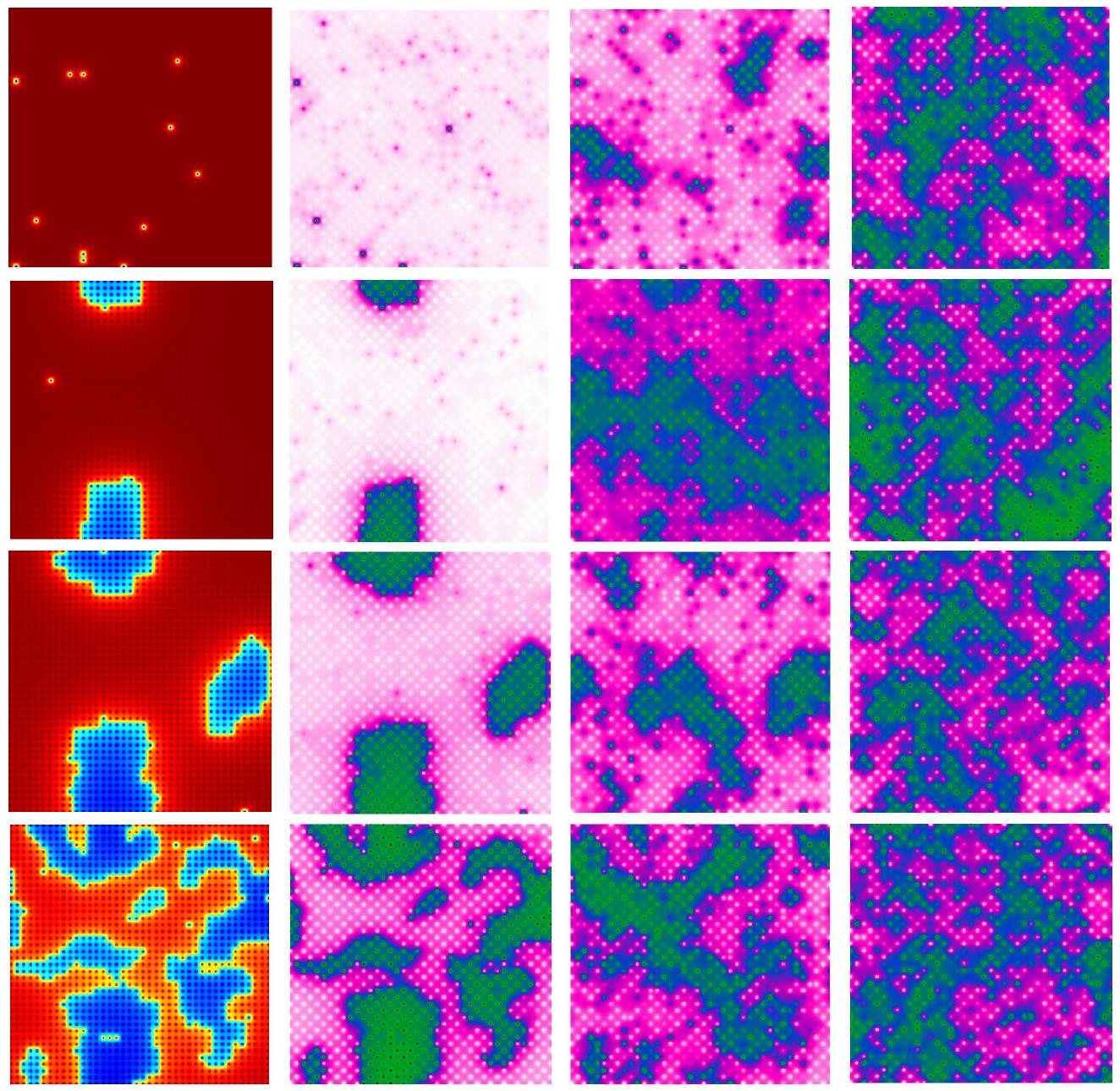} }
\caption{Colour online: Temperature dependence of short range magnetic correlations.
The left column shows ASD configurations, one from each disorder family.
$S=0.98,~0.76,~0.50$ and $0.08$, top to bottom. The 2nd, 3rd and 4th panel
along each row is a map of a spin overlap factor, $g_i$, in a Monte Carlo 
snapshot.
$g_i = {\bf S}_0.{\bf S}_i$, where ${\bf S}_0$ is the left lower corner spin
in the lattice. The temperatures are $T/t =0.03,~0.05~0.07$.
}
\end{figure}

For a given $\{\eta_i\}$
configuration 
we need to solve for the magnetic
and electronic properties.
Since the background involves strong
disorder, and the electron-spin coupling, $J$, is large, we use
an exact diagonalization based 
Monte Carlo (ED-MC) technique.
This uses the Metropolis algorithm where a spin update,
${\bf S}_i \rightarrow {\bf S}'_i$ is accepted or rejected
depending on $\Delta E/k_BT$, where $\Delta E = E ({\bf S}'_i)
-E ({\bf S}_i)$.
In principle we should diagonalise the full system every time
an update is attempted. The cost, for a large system, is 
prohibitive, so we employ a method \cite{tca}
where we diagonalize 
a cluster Hamiltonian built around the update site.

Electronic properties are calculated after 
equilibration by diagonalizing the full system.
The optical conductivity is calculated via the 
the Kubo formulation. The `dc conductivity'
is the low frequency average \cite{cond-calc},   
$ \sigma_{dc} = 
(1/{\Delta \omega})
\int_{0}^{\Delta\omega}\sigma(\omega)d\omega
$, where $\sigma(\omega)$ is thermal and disorder averaged, and 
$\Delta \omega \sim 0.05t$.
Our `dc resistivity' is the inverse of this $\sigma_{dc}$.
The spin resolved density of states $D_{\sigma}(\omega)$ 
is calculated from the single particle Green's function.

\section{Earlier work}

The theoretical effort till now 
\cite{asd-theor1,asd-theor2}
has focused on {\it uncorrelated}
antisite disorder, and quantified the impact of such ASD on
magnetic properties. One of them \cite{asd-theor1}
is based on a variational
scheme in large 3D systems, and quantifies the doping and
antisite concentration dependence of magnetic properties.
The other \cite{asd-theor2} uses a classical spin model for
magnetism, and studies the critical properties.
These calculations set the reference for magnetic properties
but have the obvious limitation that:
$(i)$~they use randomly located antisites, 
$(ii)$~they do not clarify the electronic properties, and
$(iii)$~the estimate of 
localisation effects that arise from
structural/magnetic domains, and the possible MR, remains
unexplored.

\section{Results}
\subsection{ASD configurations}
The ASD configurations can be readily generated through a Monte Carlo on
the lattice gas model. Below an ordering temperature $T_{ord} \sim 0.7V$
(in 2D) the model exhibits long range B-B' order {\it provided one anneals
long enough}. We quench the system from high temperature (random B, B')
to $T_{ann} < T_{ord}$, 
but deliberately anneal it for a short time, preventing
equilibriation. The details have been discussed \cite{ps-pm-asd}
in an earlier paper.
The four families we chose, see
Fig.1, have a fraction of mislocated sites:
$x = 0.01,~0.12,~0.25,~0.46$. The patterns, however, are strongly
correlated. Even in the most disordered samples (lowest row in Fig.1)
where the likelihood of any site being B or B' is $\sim 0.5$, if a 
site is B, say, there is a high likelihood that its neighbours will be
B' (and vice versa). Following recent experimental work
\cite{dd-asd-dom}, we calculate
the probability $p$ of having B-B' nearest neighbours as a measure of
short range correlation.
In a perfectly ordered sample this would be $1$ (the B, B' alternate)
while in a completely disordered sample this is $0.5$. For an
uncorrelated B, B' distribution this is $p_{uncorr}  = x^2 + (1-x)^2 =
(1/2) (1 + S^2)$. If we only knew $S$, these would lead to
$p_{uncorr} = 0.98,~0.79,~0.63,~0.50$ for the
$x$ that we have used.  
The values that actually emerge from analysing 
our correlated patterns, are 
$p_{corr} \sim 0.98,~0.97,~0.95,~0.86$. 
Even the most disordered samples have a high degree of short range order.
A Lorenztian fit to the B-B' structure factor, of the form
$S_{BB'}({\bf q}) \sim
{\xi}^{-1}/((q_x - \pi)^2 + (q_y - \pi)^2 + \xi^{-2})$, yields
 $\xi \sim 6.6,~5.9,~4.8,~3.6$.
\begin{figure}[t]
\centerline{
\includegraphics[width=4.4cm,height=4.4cm,angle=0]{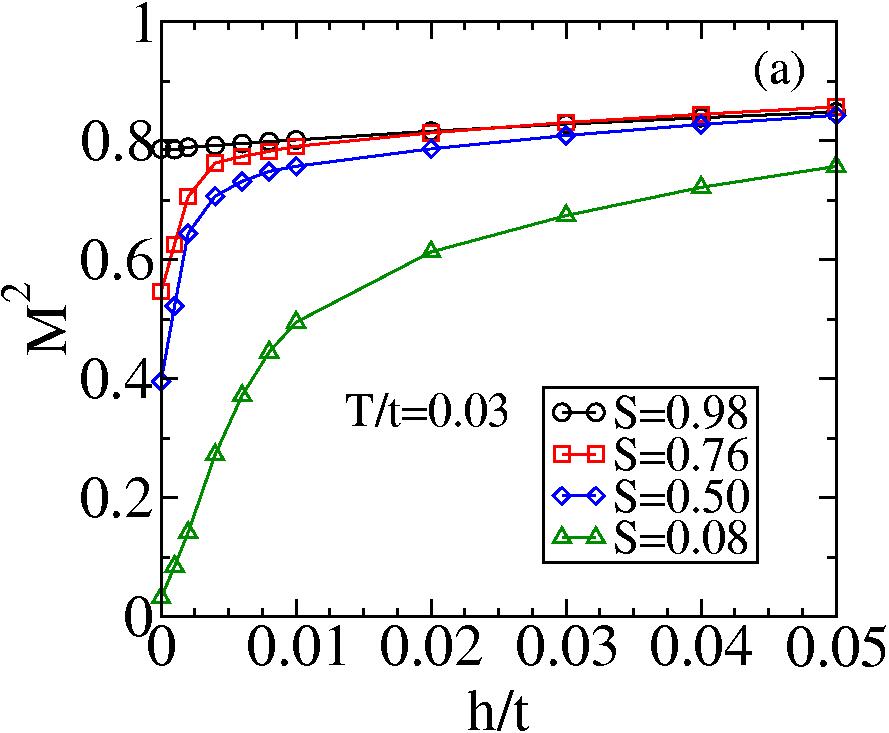}
\includegraphics[width=4.4cm,height=4.4cm,angle=0]{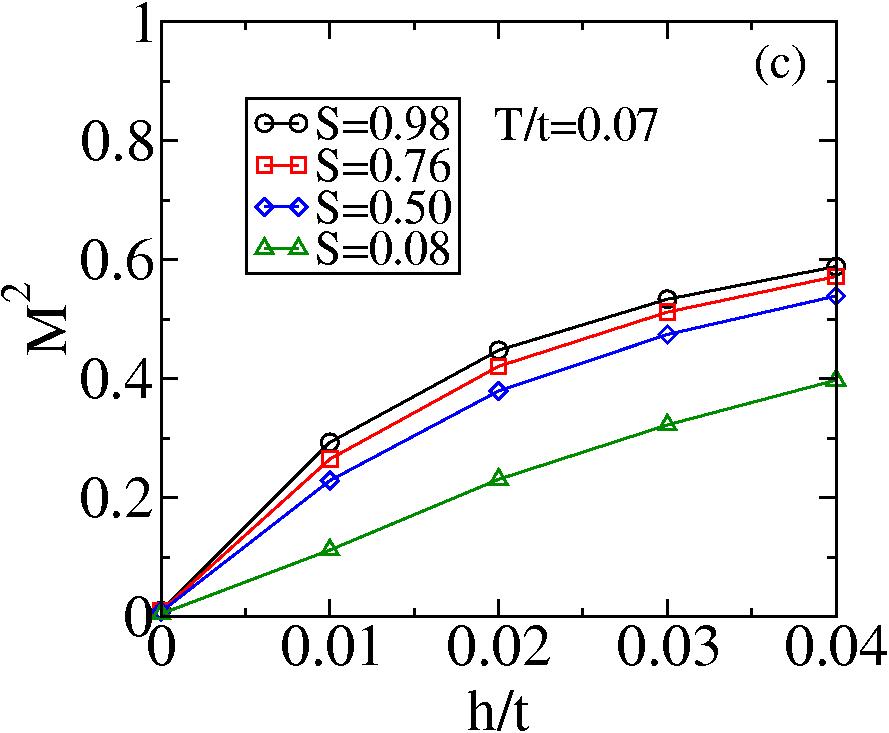}}
\vspace{.2cm}
\centerline{
\includegraphics[width=4.4cm,height=4.4cm,angle=0]{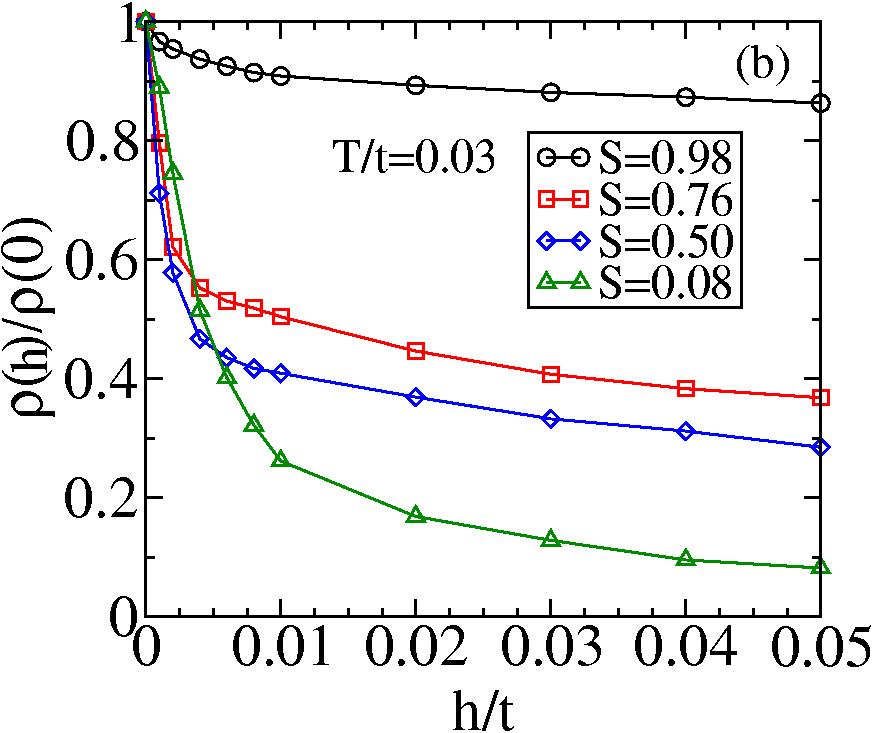}
\includegraphics[width=4.4cm,height=4.4cm,angle=0]{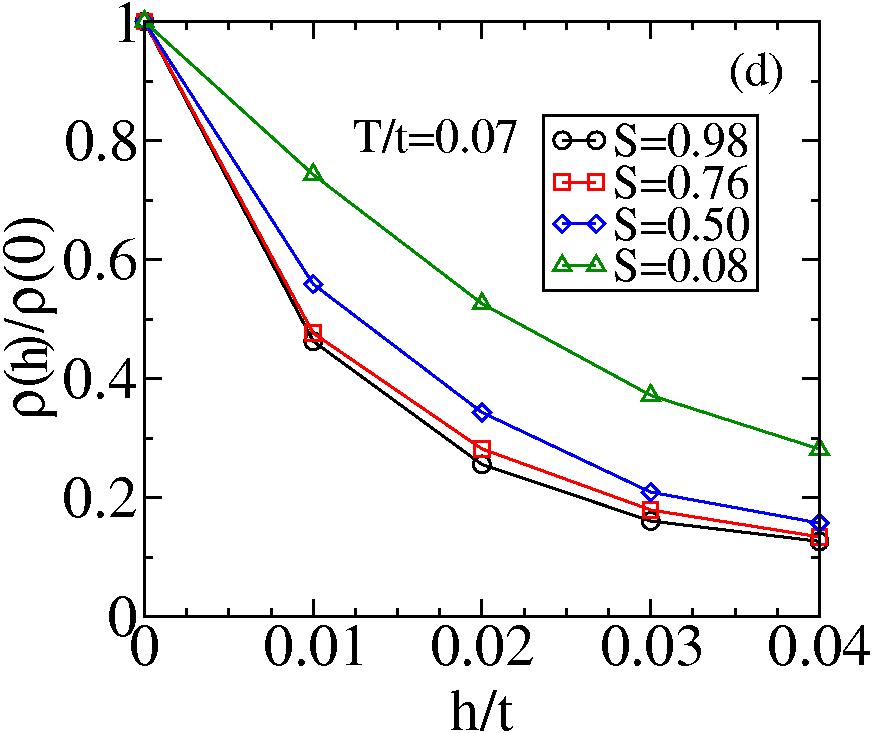}}
\caption{Colour online: (a)\&(c)~Field dependence of magnetisation, $M^2$,
(b)\&(d)~field dependence of the resistivity, normalised to $h=0$.
The left panels are at `low temperature',
 $T/t = 0.03$, the right at `high temperature', $T/t=0.07$.}
\end{figure}

\begin{figure}[t]
\centerline{
\includegraphics[width=6.6cm,height=8.8cm,angle=0]{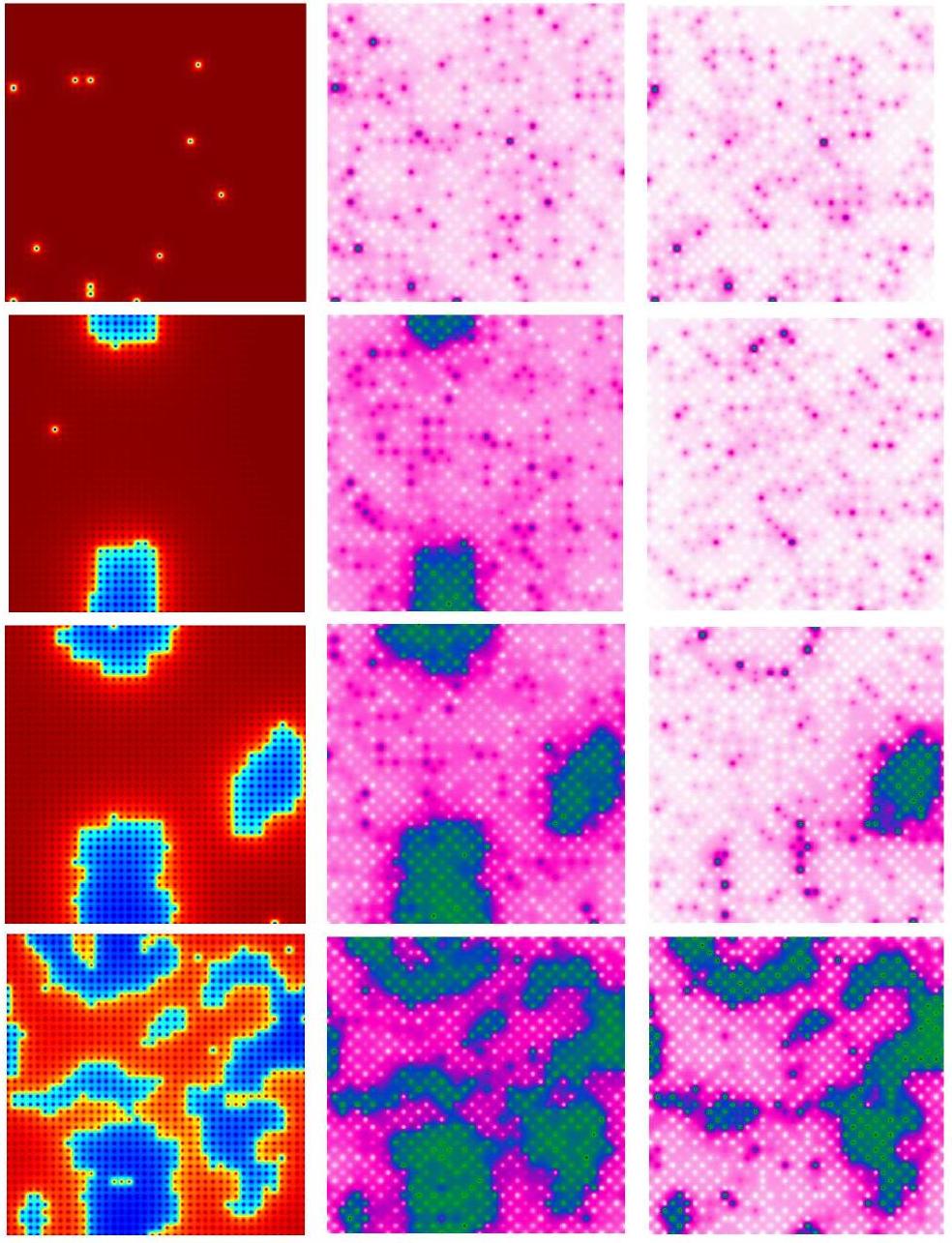}}
\caption{Colour online: Field dependence of magnetic spatial correlations.
We show the usual $g_i = {\bf S}_0.{\bf S}_i$, defined earlier. The left
column shows the ASD domains, the central column shows $g_i$ at $h/t=0$,
and the right column is for $h/t =0.001$. The temperature is $T/t =0.03$.
}
\end{figure}

\subsection{Disorder dependence at $T=0$}
Let us examine the effect of the ASD on the magnetic properties.
Suppose the fraction of  mislocated B, B' sites is  $x$, and
the structure is organised into domains such that the ratio of
`perimeter' to `bulk' sites is small. The  AF coupling between
adjoining domains would polarise them antiparallel, and the 
net moment at $T=0,~h=0$ would be proportional to the volume 
difference of up and down domains. We should have 
$M(T=0, h=0) \sim (1 - x) - x = 1 - 2x = S$ so $M^2 = S^2$. Given
our $S$, these are $0.96,~0.58,~0.25,~0.01$ in almost
perfect correspondence with the $T \rightarrow 0$ values in Fig.2.(a).
The elaborate calculation arrives at an obvious answer.
The onset temperature for magnetic order seems to be insensitive to
the ASD, {\it i.e.}, the intra-domain order 
sets in at $T \sim $ the bulk $T_c$.
While our answer for the suppression of magnetisation is $M \sim 1 -2x$,
a 3D calculation, with uncorrelated disorder, had
found \cite{asd-theor1} $M \sim 1 - 1.9x$.

\begin{figure}[t]
\centerline{
\includegraphics[width=5.0cm,height=5.0cm,angle=0]{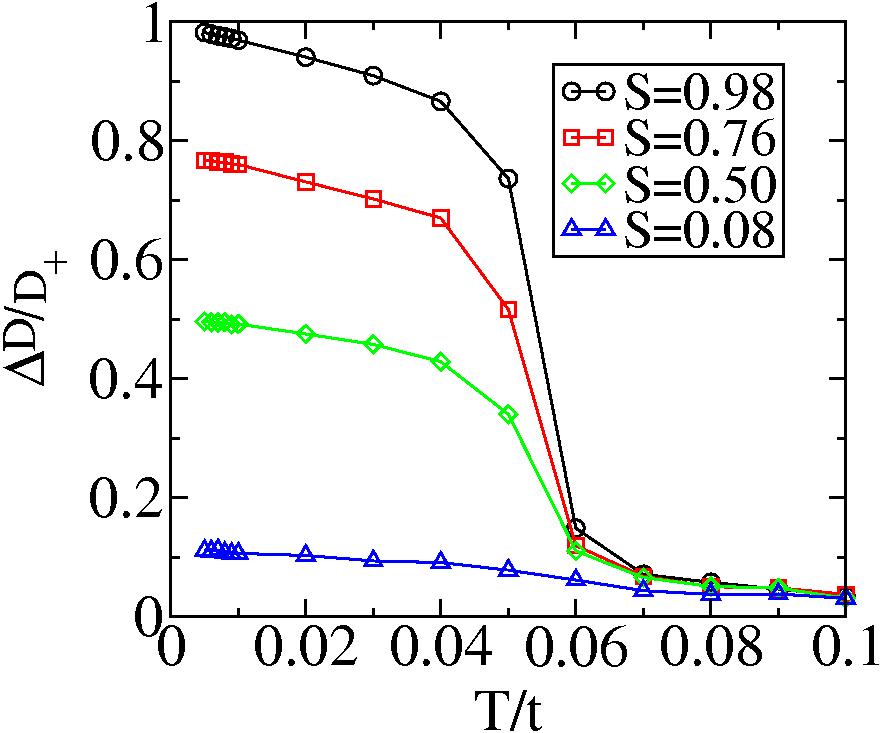}}
\caption{Colour online: Half-metallicity, estimated as 
$(D_{\uparrow}(\epsilon_F) - D_{\downarrow}(\epsilon_F))/
(D_{\uparrow}(\epsilon_F) + D_{\downarrow}(\epsilon_F))$, for 
varying temperature and different degrees of antisite disorder.}
\end{figure}

The transport in these background is shown in Fig.2.(b).
From weak to intermediate ASD the temperature dependence of
$\rho(T)$ remains similar,
with a sharp drop near $T_c$. The only effect of increasing ASD is
an increase in residual resistivity.
It is as if there is a temperature independent `structural' scattering
that gets added to the $T$ dependent magnetic scattering. At large 
ASD, however, this correspondence breaks down, the
$T=0$ `resistivity' is very large (and grows with growing system
size) and $d\rho/dT <0$. There seems to be a metal-insulator
transition between $S=0.50$ and $S=0.08$ 

To create an understanding of this let us focus on
$T=0$, where the magnetic configuration is simple (collinear).
The $\downarrow$ spin electrons inhabit the ${\bf {\uparrow}}$ core 
spin domains, and {\it vice versa}. The conductance arises from
the inter-penetrating parallel channels for up and down spin
electrons. One could call it ``complementary percolation''. 
Let us identify up electrons with the `majority' phase, and 
down with the `minority' domains.
The net conductivity $\sigma_{tot}(S) = \sigma_{maj}(S) + 
\sigma_{min}(S)$. While this reduces monotonically with
reducing $S$ in our data, Fig.2.(b), the conductivity also depends
on $p_{corr}$. Even at $S=0$, one could increase $\sigma_{tot}$
systematically by increasing $p_{corr}$, {\it i.e}, reducing
the fragmentation of the conduction paths. Weak localisation  
effects, {\it etc}, in two dimensions 
could show up at much longer lengthscales,
but at a given cutoff size the trend above would survive.

\subsection{Temperature dependence:} 
The ASD configuration is
$T$ independent so the primary sources of $T$ dependence on
transport are (i)~the weakening of AF locking across the domain
boundaries, and (ii)~fluctuations about the FM state {\it within
a domain}. The first effect enhances the conductivity, while the
second serves as a source of scattering. Their relative importance
depends on $\sigma(T=0)$.
For weak disorder (large $S$) one is far from the percolation
threshold and the decrease in $\sigma$ due to intra-domain magnetic 
scattering
is larger than  the enhancement from inter-domain tunneling.
However, by the time $S=0.50$ there is already a weak upturn 
in $\rho$ as $T \rightarrow 0$, the intra-domain effect is visible,
and this becomes the dominant effect as $S \rightarrow 0$. 

An analysis of the spatial spin-spin correlations illustrates
the AF locking of domains at low temperature and how this
weakens with increasing $T$. The first column in Fig.3 reproduces
one set of ASD configurations from Fig.1 (the first column).
The next three columns show the
the magnetic overlap $g_i = {\bf S}_0.{\bf S}_i$, where ${\bf S}_0$
is the lower left corner spin in each configuration, for MC
snapshots at $T/t=0.03,~0.05,~0.07$. These pictures would
correspond to  `magnetic domains' if the patterns survived thermal
averaging.
The low temperature snapshots correspond closely to the ASD pattern.
The antiphase boundary (APB) and the magnetic domain wall (MDW)
pattern coincide at $T/t = 0.03$. At $T/t=0.05$, however,
close to the bulk $T_c$, there is no correlation between the
APB and the $g_i$ pattern. 
There is {\it significant core spin overlap} across the boundary,
and large fluctuation, overall, in spin orientation. This bears
out the transport mechanism we suggested in the preceding
paragraph.

\subsection{Field dependence}
The field dependence of magnetisation and resistivity is shown
in Fig.4., at relatively low temperature, $T/t=0.03$, in
(a)-(b), and high temperature, $T/t=0.07$, in (c)-(d).
Three energies play out when $h \neq 0$: (i)~the bulk Zeeman cost of
the `minority' domains, $\sim h V_{min}$, where
$V_{min}$ is the volume of the minority phase, (ii)~the
interfacial AF energy, $\sim J_{AF} V (1 - p_{corr}) $, where
$1 - p_{corr}$ is the fraction of AF bonds on the lattice, and
(iii)~the {\it gain in electronic kinetic energy} on removal (or rotation)
of MDW's. (i) and (iii) prefer domain alignment while (ii) prefers
to retain domain walls. In a `spin only' model (iii) would be
absent. This delocalisation energy gain serves to reduce the field
at which domain rotation can occur.

At low $T$, Fig.4.(a)-(b),
the ordered samples have a high degree of magnetic
order, so the field induced 
increase in $M$, and the decrease $\Delta \rho/\rho(0)$,
where $\rho(0) = \rho(h=0)$ and $\Delta \rho = \rho(0)-\rho(h)$,
is quite small.
The low $T$ low field response is, however, dramatic for low $S$ samples.
These samples have 
$M(h=0) \sim 0$, and a large $\rho(0)$ due to the 
fragmented (spin selective) conduction path.
A field as small as $h/t \sim 0.001$ leads to $M^2 \sim 0.1$, so
$M \sim 0.3$. The corresponding impact on spin correlations is shown
in the lowest row in Fig.5, where the MDW pattern is strongly
affected by $h$. While the domain rotation effect is visible 
both for $S=0.50$ and $S=0.08$, the less disordered sample had
a larger conductivity at $h=0$ so the fractional change is
much larger for $S=0.08$. 

At high temperature, Fig.4.(c)-(d), the domains
cease to exist and conductance gain from  domain rotation 
is irrelevant. In the large $S$ samples there are few AF
links so the applied field just suppresses the magnetic
fluctuations leading to large $\Delta \rho/\rho(0)$. 
In the most  disordered samples there are $ (1-p_{corr})/2
\sim 7\%$ of AF bonds. Although there are no domains, these
act as a source of scattering. 
The gain in conductivity is slower in the disordered samples
compared to the more ordered ones.

\subsection{Half metallicity}
These systems are unusual because at $T=0$ 
within each domain the
conduction electron has only one spin polarisation but
averaged over the system both $\uparrow$ and $\downarrow$
electrons have density of states at $\epsilon_F$. A
local probe, with probe area $< \xi^2$, will allow
only spin polarised tunneling, while a probe averaging over
domains will see both $D_{\uparrow}(\epsilon_F)$ and
$D_{\downarrow}(\epsilon_F)$. Fig.6 shows 
$(D_{\uparrow}(\epsilon_F) - D_{\downarrow}(\epsilon_F))/ 
(D_{\uparrow}(\epsilon_F) + D_{\downarrow}(\epsilon_F))$ as
a measure of half-metallicity. It is unity only in the
absence of ASD and at $T=0$, and in general has a 
behaviour that broadly mimics the behaviour of the
core spin magnetisation, Fig.2.(a).

\section{Discussion}
There are two issues we want to touch upon, to relate our work
to real double perovskites. (a).~The role of dimensionality:
it is well known that localisation effects are stronger in 2D
compared to 3D, so we ran this entire calculation on a $12^3$
system to check out the trends in transport. In particular we
confirmed that there indeed is a sharp rise in the residual
resistivity (although possibly no insulating phase) with
increasing ASD. The low temperature upturn in $\rho(T)$ is
present, but weaker, even in 3D. (b).~Role of grain
boundaries: in the absence of a chemical characterisation
of the grain boundary material, and an electronic model
for the GB, it is hard to construct a comprehensive theory.
However, since grain size, $l_G >> \xi$, it should be
possible to study the role of APB's and MDW's via non
contact probes that focus on a single grain.

\section{Conclusion}

We have studied a double perovskite model
on antisite disordered backgrounds with a high degree of
short range correlation. 
In this situation, the antiphase boundaries coincide
with the $T=0$  magnetic domain walls.
Growing ASD reduces the low field
magnetization, destroys the half-metallicity, and leads to 
a low temperature metal-insulator transition.  
While these are disadvantages, we also note that the
ferromagnetic $T_c$ is only weakly affected by moderate ASD
and the low field magnetoresistance is dramatically
enhanced by disorder. 
Our real space results allow an interpretation of these in 
terms of the domain pattern, the effective exchange, and the
short range magnetic correlations. They are also
consistent with explicit spatial imagery from recent experiments.
The `intra-grain' effects highlighted here would be 
directly relevant to single crystals, and define the
starting point for a 
transport theory of the polycrystalline double perovskites.

{\it Acknowledgements:} 
We acknowledge use of the Beowulf Cluster at HRI and discussions with
G. V. Pai, P. Sanyal, D. D. Sarma, and R. Tiwari. 
PM acknowledges support from a DAE-SRC Outstanding Research Investigator
Award, and the DST India through the Indo-EU ATHENA project.


\begin{thebibliography}{99}
\bibitem{dp-rev} For reviews, see
D. D. Sarma, Current Op. Solid St. Mat. Sci.,{\bf 5}, 261 (2001),
D. Serrate, J. M. de Teresa and M. R. Ibarra,
J. Phys. Cond. Matt. {\bf 19}, 023201 (2007).
\bibitem{nat-kob} K.-I. Kobayashi, T. Kimura, H. Sawada,
K. Terakura and Y. Tokura, Nature {\bf 395}, 677 (1998).
\bibitem{tom-cryst}
Y. Tomioka, T. Okuda, Y. Okimoto, R. Kumai, K.-I. Kobayashi,
and Y. Tokura,
Phys. Rev. B {\bf 61}, 422 (2000).
\bibitem{ins-pap} K.-I. Kobayashi, T. Okuda, Y. Tomioka, T. Kimura
and Y. Tokura,
Jl  Magn and Magn Matls, 218, 17 (2000).
\bibitem{lanimn-dd}
H. Das, U. V. Waghmare, T. Saha-Dasgupta, and D. D. Sarma,
Phys. Rev. Lett. 100, 186402 (2008).
\bibitem{theor-abin-dd}
D. D. Sarma, P. Mahadevan, T. Saha-Dasgupta, S. Ray, and A. Kumar,
Phys. Rev. Lett. {\bf 85}, 2549 (2000).
\bibitem{theor-millis}
A. Chattopadhyay and A. J. Millis,
Phys. Rev. B {\bf 64}, 024424 (2001).
\bibitem{ps-pm-scr} P. Sanyal and
P. Majumdar, Phys. Rev. B {\bf 80}, 054411 (2009).
\bibitem{garcia-hern}
M. Garcia-Hernandez, J. L. Martinez, M. J. Martinez-Lope,
M. T. Casais, and J. A. Alonso,
Phys. Rev. Lett. {\bf 86}, 2443 (2001).
\bibitem{huang-asd1} Y. H. Huang, M. Karppinen, H. Yamauchi,
and J. B. Goodenough,
Phys. Rev. B {\bf 73}, 104408 (2006).
\bibitem{huang-asd2}
Y. H.  Huang, H. Yamauchi, and M. Karppinen,
Phys. Rev. B {\bf 74}, 174418 (2006).
\bibitem{nav-asd1}
J. Navarro, L. Balcells, F. Sandiumenge, M. Bibes, A. Roig, B. Martınez and
J. Fontcuberta,
J. Phys. Cond. Matt {\bf 13}, 8481 (2001).
\bibitem{nav-asd2}
J. Navarro, J. Nogues, J. S. Munoz, and J. Fontcuberta,
Phys. Rev. B {\bf 67}, 174416 (2003).
\bibitem{dd-gb-mr} D. D. Sarma, S. Ray, K. Tanaka, M. Kobayashi,
A. Fujimori, P. Sanyal, H. R. Krishnamurthy and C. Dasgupta,
Phys. Rev. Lett.,{\bf 98}, 157205 (2007).
\bibitem{asd-theor1}
J. L. Alonso, L. A. Fernandez, F. Guinea, F. Lesmes, and
V. Martin-Mayor,
Phys. Rev. B {\bf 67}, 214423 (2003).
\bibitem{asd-theor2}
C. Frontera and J. Fontcuberta,
Phys. Rev. B {\bf 69}, 014406 (2004).
\bibitem{asaka-asd-dom} T. Asaka, X. Z. Yu, Y. Tomioka, Y. Kaneko, T. Nagai,
K. Kimoto, K. Ishizuka, Y. Tokura, and Y. Matsui,
Phys Rev B {\bf 75}, 184440 (2007).
\bibitem{dd-asd-dom}
C. Meneghini, Sugata Ray, F. Liscio, F. Bardelli, S. Mobilio, and D. D. Sarma,
Phys. Rev. Lett. {\bf 103}, 046403 (2009).
\bibitem{niebel-asd}
D. Niebieskikwiat, F. Prado, A. Caneiro, and R. D. Sanchez,
Phys. Rev. B {\bf 70}, 132412 (2004).
\bibitem{ps-pm-asd} P. Sanyal, S. Tarat and P. Majumdar,
Eur. Phys. J. {\bf B  65}, 39 (2008).
\bibitem{delta-ref}
In SFMO $(\epsilon_B - J/2) -\epsilon_{B'} \sim 5t$, and we have
checked that this B, B'  energy mismatch mainly serves to
enhance the resistivity without affecting the trends we observe.
\bibitem{tca} S. Kumar and P.Majumdar,
Eur. Phys. J. {\bf B 50}, 571-579 (2006).
\bibitem{cond-calc}S. Kumar and P. Majumdar,
Eur. Phys. J. {\bf B 46}, 237 (2005).
\end{thebibliography}
\end{document}